\documentstyle[psfig,12pt]{article}
\def\TL{\hfil$\displaystyle{##}$}
\def\TR{$\displaystyle{{}##}$\hfil}

\def\comment#1{}
\def\fixit#1{}

\def\overleftrightarrow#1{\vbox{\ialign{##\crcr
     $\leftrightarrow$\crcr\noalign{\kern-0pt\nointerlineskip}
     $\hfil\displaystyle{#1}\hfil$\crcr}}}

\def\lsim{\mathrel{\mathstrut\smash{\ooalign{\raise2.5pt\hbox{$<$}\cr\lower2.5pt\hbox{$\sim$}}}}}
\def\gsim{\mathrel{\mathstrut\smash{\ooalign{\raise2.5pt\hbox{$>$}\cr\lower2.5pt\hbox{$\sim$}}}}}
\def\sqr#1#2{{\vcenter{\vbox{\hrule height.#2pt
         \hbox{\vrule width.#2pt height#1pt \kern#1pt
            \vrule width.#2pt}
         \hrule height.#2pt}}}}
\def\square{\mathop{\mathchoice\sqr56\sqr56\sqr{3.75}4\sqr34\,}\nolimits}
\def\href#1#2{#2}  

\def\lbldef#1#2{\expandafter\gdef\csname #1\endcsname {#2}}
\def\eqn#1#2{\lbldef{#1}{(\ref{#1})}%
\begin{equation} #2 \label{#1} \end{equation}}
\def\eqalign#1{\vcenter{\openup1\jot
    \halign{\strut\span\TL & \span\TR\cr #1 \cr
   }}}

\def\comment#1{  \begin{raggedright}{\tt [#1]}\end{raggedright}}
\begin{document}
\baselineskip=15.5pt
\pagestyle{plain}
\setcounter{page}{1}
%--------+---------+---------+---------+---------+---------+---------+

%Title page

\begin{titlepage}

\begin{flushright}
PUPT-1973 \\
hep-th/0012155
\end{flushright}
\vfil

\begin{center}
{\Large D-brane Dynamics and the Quantum Hall Effect}
\end{center}

\vfil
\begin{center}
{\large  Steven S. Gubser and Mukund Rangamani}
\end{center}

\begin{center}
Joseph Henry Laboratories, Princeton University, Princeton,
NJ 08544.
\end{center}
\vfil

\begin{center}
{\large Abstract}
\end{center}

\noindent 
 We study the recently proposed D-brane configuration \cite{bbst}
modeling the quantum Hall effect, focusing on the nature of the
interactions between the charged particles.  Our analysis indicates
that the interaction is repulsive, which it should be for the ground
state of the system to behave as a quantum Hall liquid.  The strength
of interactions varies inversely with the filling fraction, leading us
to conclude that a Wigner crystal is the ground state at small $\nu$.
For larger rational $\nu$ (still less than unity), it is reasonable to
expect a fractional quantum Hall ground state.

\vfil
\begin{flushleft}
December 2000
\end{flushleft}
\end{titlepage}
\newpage
%--------+---------+---------+---------+---------+---------+---------+
%Body 
\section{Introduction}
\label{Intro}

\begin{figure}[h]
   \centerline{\psfig{figure=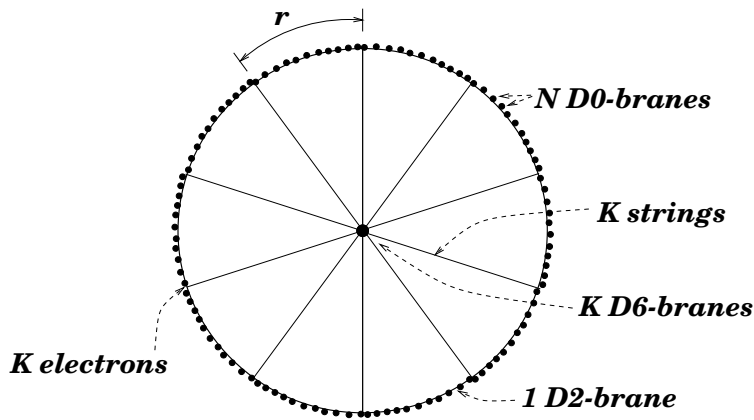,width=4in}}
   \caption{The ``quantum Hall soliton'' of \cite{bbst}.  The
D6-branes are viewed end-on: they should be thought of as projecting
out of the page in six orthogonal directions.}
   \label{figA}
  \end{figure}

In \cite{bbst} it was conjectured that a specific assembly of D-branes
and fundamental strings would have a low-energy dynamics similar to
systems displaying the fractional quantum Hall effect (FQHE).
Specifically, a D2-brane in the shape of ${\bf S}^2 \times {\bf R}$ is
placed around $K$ flat D6-branes, so that the spatial directions of
the different branes are orthogonal.  The radius $\rho$ transverse to
the D6-brane is the one spatial direction perpendicular to both brane
world-volumes.  Because of a topological constraint, $K$ fundamental
strings stretch from the D6-branes to the D2-brane.  The ends of these
strings carry electric charge in the $U(1)$ gauge theory on the
D2-brane world-volume.  A large number of D0-branes are bound to the
D2-brane, representing strong magnetic flux in this $U(1)$ gauge
theory.  See figure~\ref{figA}.

The string ends on the D2-brane are the ``electrons'' (and will
hereafter be referred to as such), and the D0-branes are the flux
quanta.  The infrared dynamics is supposed to involve nearly rigid
motion of the strings, and possibly a binding of D0-branes to the
strings as a manifestation of binding flux quanta to
electrons.\footnote{D0-branes stuck to a D2-brane would ordinarily be
``dissolved,'' even in their classical description, to produce a
uniform D0-charge (or magnetic field) on the D2-brane.  It is not
clear to us that the binding of individual D0-branes to strings is to
be taken literally as a stringy analog of the binding of flux quanta,
which is better described as a change of variables than as a
localization of the physical magnetic field.  Optimistically, some
appropriate change of variables in the D-brane setup would produce
quasi-D0-branes which can change the statistics of the electrons but
don't carry the magnetic field.  We thank S.~Sondhi for a discussion
on this point.}  Clearly this system exhibits features in common with
quantum Hall systems.  However it is known that putative quantum Hall
systems exhibit a variety of phases, including the Wigner crystal and
stripe phases (see \cite{Sondhi,Girvin} for pedagogical reviews and
references to the extensive condensed matter literature).  It is the
purpose of this note to inquire whether a fractional quantum Hall
liquid is ever the ground state of the system.  In general, this is a
hard question which can be answered definitively only by diagonalizing
the complete Hamiltonian (including inter-electron interactions).  We
won't do this; but we will compute the force between electrons and
find (modulo a plausible technical assumption) that it is repulsive.
This is good because attractive interactions would inevitably lead to
a clumping instability and no quantum Hall behavior.  The
characteristic energy scale of the interactions is comparable to the
cyclotron gap, which is evidence that a quantum Hall liquid is at
least not parametrically disadvantaged when compared to a Wigner
crystal.  Thus, our results lead us to be cautiously optimistic that a
quantum Hall ground state exists at least for some filling fractions.

It is conceivable that some modification of the proposal of
\cite{bbst} would in fact parametrically favor a quantum Hall ground
state in the thermodynamic limit, $N \to \infty$.  The main desiderata
are to weaken the inter-electron force and/or raise the cyclotron gap.
If the goal is to have quantized transverse conductance, impurities
are essential.  For a clean subject like string theory, this may be
the hardest part.

In section~\ref{CondMat} we will briefly review some of the salient
points of quantum Hall physics relevant to our analysis.  In
section~\ref{Force} we will compute the inter-electron force and show
that it is repulsive.  This is a slightly delicate computation because
the electrons almost enjoy a BPS no-force condition.  Only finite
volume effects break supersymmetry and thereby alter the BPS
condition.  The near-cancellation of inter-electron forces arising
from scalar and gauge boson exchange can be viewed heuristically as an
odd type of screening of the electrostatic repulsions which becomes
more and more complete the closer the electrons get to one another.
In section~\ref{WhichPhase} we demonstrate that the repulsive
interactions are of the same order of magnitude as the cyclotron gap.

While this paper was in preparation, \cite{bena} appeared, discussing
possible instabilities of the D-brane system set up in \cite{bbst}.
It was shown that, if there is no binding of flux quanta, there are
instabilities in the $\ell=1$ and $\ell=2$ partial waves on the ${\bf
S}^2$; but if flux quanta do bind, there is no instability.  This work
is in a sense orthogonal to ours, since we focus on the inter-electron
force and regard the binding of flux quanta as a derivative effect.
Considerations similar to \cite{bena} may affect the stability of the
Wigner crystal state toward long-wavelength fluctuations.

\section{Some aspects of quantum Hall physics}
\label{CondMat}

Knowing the sign on the force between two string ends is important
because it affects whether a quantum Hall state will form.  The
dynamical criteria for formation of quantum Hall states are, roughly, 
  \begin{itemize}
   \item[1)] There should be a repulsive force between charges.
   \item[2)] The typical energy of these repulsive interactions should be
less than the cyclotron gap, $\omega_c = eB/m_{\rm electron}c$.
   \item[3)] Charges should be crowded to within distances shorter than
the uncertainties in their respective positions.
  \end{itemize}
 If 1) fails, then the charges tend to clump together.  No quantum
Hall state will form.  The binding of flux quanta in real FQHE systems
occurs to lower the repulsive Coulomb energy: the high power of
$z_i-z_j$ in Laughlin's wave-function keeps the charges apart.
Repulsive interactions between charges are the {\it sine qua non} of
the fractional quantum Hall effect.

If 1) holds but 2) or 3) fail, then a Wigner crystal is generally
preferred over the quantum Hall state.  In fact, the Wigner crystal is
a much more generic state of matter for variants of the
two-dimensional electron gas.  In \cite{bbst}, it was argued that a
clean infrared limit existed where a quantum Hall ground state might
be seen, but the arguments depended on having small filling fraction.
In real two-dimensional electron gas systems with filling fraction
below about $\nu = 1/7$, the conductance plateaux disappear and the
ground state becomes a Wigner crystal (though without long-range
order, since it's in two dimensions).  We expect similar behavior in
the D-brane system, but for a different reason than in real quantum
Hall systems: as we will show in section~\ref{WhichPhase}, criterion
2) fails when $\nu$ is very small.

An issue which was left open in \cite{bbst} is whether the electrons
behave as fermions or bosons on the D2-brane.  It seems most plausible
that the electrons behave as bosons in their ground state: as a whole,
the D2-D6 strings are fermions in their ground state, but the $K$
string ends on the D6-branes need to be assembled into a gauge singlet
of the $U(K)$ gauge theory with a $\epsilon_{i_1 \ldots i_K}$ tensor.
As discussed in \cite{bbst}, an antisymmetric spatial wavefunction on
the D6-brane world-volume would change the statistics of the electrons
back to fermions.  Bosons can form fractional quantum Hall states at
even filling fractions: the Laughlin wave-function would involve even
powers of $z_i-z_j$.  Little of our analysis will rely on the
statistics of the electrons in their ground state.

Finally, it is perhaps worth recalling the value of dirt in the
quantum Hall effect.  By ``dirt'' we mean quenched impurities. In the
absence of dirt, one might invoke the Lorentz symmetry of the system
to infer that the Hall conductance varies inversely with the magnetic
field, the slope being proportional to the density of the electrons.
At certain rational filling fractions one might still expect that the
Laughlin wave-function is the ground state.  Near such a ground state,
the quasi-particle excitations will give rise to finite $\sigma_{xx}$,
and $\sigma_{xy}$ will vary with the filling fraction, so the
characteristic plateaux will be absent from the conductance profile.
When there is dirt, these quasi-particles localize, so $\sigma_{xx} =
0$ provided the Fermi level of the quasi-particle excitations lies
within the energy gap and the number of quasi-particles is
insufficient to drive the system into the next plateau.  The
transverse conductance, $\sigma_{xy}$, receives contributions only
from boundary states, and is quantized.  It is somewhat analogous to
the $\theta$-angle in QCD, and it is independent of bulk
characteristics like the geometry of the sample.  To summarize, in
totally clean samples like the ones we will consider the transverse
conductivity profile won't exhibit the familiar plateaux; but the
quantum Hall ground state can still prevail at isolated filling
fractions.

\section{The inter-electron force}
\label{Force}

A natural description of the magnetic flux is to make the D2-brane
gauge theory non-commutative.  A second consequence of the flux is
that it introduces a Chern-Simons interaction into the gauge theory.
This is effectively a mass term for the photon.  The gauge theory also
includes scalars corresponding to the transverse fluctuations of the
D2-brane.  These scalars couple to the string ends.  For radially
directed fundamental strings, the scalar corresponding to radial
fluctuations is the only one that couples to the string ends.  This
scalar is massive because the radius of the D0-D2 bound state is
stable to spherical perturbations.

The force between two fundamental strings arises from two sources.
First, there is an attractive force from bulk effects.  Unless the
strings run parallel ({\it i.e.}\ unless they are coincident), the
attraction from graviton and dilaton exchange overcomes the repulsion
from $B_{\mu\nu}$ exchange, because the strings are at angles.
Second, there is a force from the dynamics of the D2-brane world
volume theory.  We will argue that this force is stronger in the large
$N$ limit.  It is not so obvious {\it a priori} whether it is
attractive or repulsive.  The photon on the D2-brane world-volume
induces a repulsive force, but the radial scalar induces an attractive
force.  The coupling constant for these two forces is the same, and
they would cancel if it weren't for the effects of the D6-brane and
the curvature of the D2-brane world-volume.  So the question comes
down to whether the photon or the scalar is more massive.

The total low-energy effective action on the D2-brane world-volume, in
$+$$-$$-$ signature, is
  \eqn{LowS}{\eqalign{
   S_{\rm eff} &= \int d^3 x \, \sqrt{g} \Bigg[ J^0 A_0
     - {1 \over 4} F_{\mu\nu}^2 +
     {1 \over 2} m_\gamma \epsilon^{\mu\nu\rho} A_\mu \partial_\nu A_\rho + 
     {1 \over 2} (\partial\phi)^2 - {1 \over 2} m_\phi^2 \phi^2  \cr
    &\qquad\quad{} + \hbox{(fermions)} + 
      \hbox{(interactions)} \Bigg] 
    + \sum_{i=1}^K Q \int_{\gamma_i} (A + \phi d\tau) \,,
  }}
 where $Q$ is the charge of a string end.  There is also a uniform
background charge $J^0$ which ensures overall charge neutrality.  The
terms in the second line of \LowS\ indicate the couplings of the
string ends to the gauge field and to the radial scalar $\phi$.  We
have chosen to suppress the non-commutativity inherent in the action,
for we shall mainly be concerned with calculating propagators and to
this end modification of the action through introduction of star
products is immaterial.  The Chern-Simons and scalar masses are
  \eqn{mDef}{\eqalign{
   m_\gamma &= {1 \over 4} {1 \over (\pi N )^{1/3} l_s}  \cr
   m_\phi &=   {4\sqrt{2} \over 3} {1 \over (\pi N)^{1/3} l_s} \,,
  }}
 where $l_s = \sqrt{\alpha'}$.  We will derive \mDef\ explicitly in
sections~\ref{GaugeField} and~\ref{ScalarField}.

The remainder of this section will be devoted mainly to showing that
$m_\gamma < m_\phi$ implies that there is a net repulsion between
nearby electrons coming from D2-brane effects.

First consider a flat D2-brane with a finite density of D0-branes
bound to it and perpendicular external strings attached, the same
action would apply except with $m_\gamma=m_\phi=0$ and $J^0=0$
(assuming no D6-branes).  Such a system would be BPS, and the string
ends would exert no force on one another.  Bending the D2-brane into
the shape of an ${\bf S}^2$ breaks the supersymmetry, and we no longer
expect a no-force condition to hold in the D2-brane world-volume
theory.

Let us assume that the ${\bf S}^2$ is large, and that $K$ is also
large, and ask what force there is between two fundamental string ends
which are separated by a small angle.  For this purpose it is enough
to consider the quadratic part of the action $S_{\rm eff}$ in flat
${\bf R}^{2,1}$: we will compute only the tree-level propagators of
the gauge field and the scalar.  The theory is at weak coupling, so
this should suffice to determine the force between string ends.  The
propagator for the scalar is $W(p) = i/(p^2-m_\phi^2)$.  To obtain the
propagator for the gauge boson, we introduce a gauge fixing term and
write the action as
  \eqn{GaugeAction}{\eqalign{
   S_{\rm gauge} &= \int d^3 x \, \left[ -{1 \over 4} F_{\mu\nu}^2 + 
    {1 \over 2} m_\gamma \epsilon^{\mu\nu\rho} A_\mu \partial_\nu A_\rho + 
    {1 \over 2\xi} (\partial^\mu A_\mu)^2 \right]  \cr
   &= \int d^3 x \, {1 \over 2} 
     A_\mu \left( m_\gamma \epsilon^{\mu\nu\rho} \partial_\nu +
     g^{\mu\rho} \square \right) A_\rho  \cr
   &\equiv \int d^3 x \, {1 \over 2} A_\mu S^{\mu\rho} A_\rho \,,
  }}
 where in the second line we have gone to Feynman gauge, $\xi=1$.
Fourier transforming, one obtains
  \eqn{SForm}{\eqalign{
   S^{\mu\rho}(p) &= -g^{\mu\rho} p^2 - 
       i m_\gamma \epsilon^{\mu\nu\rho} p_\nu  \cr
     &= \pmatrix{ -p^2 & i m_\gamma p_2 & -i m_\gamma p_1  \cr
                  -i m_\gamma p_2 & p^2 & i m_\gamma p_0  \cr
                  i m_\gamma p_1 & -i m_\gamma p_0 & p^2 }\,,
  }}
 where $\epsilon^{012} = 1$.  It may seem peculiar to have imaginary
components in $S^{\mu\rho}(p)$, but the $i$'s are in the right places
to make the Minkowskian action real.  We obtain the propagator by inverting:
  \eqn{SInverse}{\eqalign{
   W_{\mu\rho}(p) &= i \left[ S^{\mu\rho} \right]^{-1}  \cr
    &= {i \over p^4 (p^2 - m_\gamma^2)}
       \pmatrix{ -p^4 + p_0^2 m_\gamma^2 & 
          i m_\gamma p_2 p^2 + p_0 p_1 m_\gamma^2 &
          -i m_\gamma p_1 p^2 + p_0 p_2 m_\gamma^2  \cr
        -i m_\gamma p_2 p^2 + p_0 p_1 m_\gamma^2 & 
          p^4 + p_1^2 m_\gamma^2 &
          -i m_\gamma p_0 p^2 + p_1 p_2 m_\gamma^2  \cr
        i m_\gamma p_1 p^2 + p_0 p_2 m_\gamma^2 & 
          i m_\gamma p_0 p^2 + p_1 p_2 m_\gamma^2 &
          p^4 + p_2^2 m_\gamma^2 }
  }}
 The potential energy arising from gauge boson exchange between two
stationary string ends is obtained by differentiating the Fourier
transform of $W_{00}(p)$ with $p$ entirely spatial.  The scalar
mediated potential energy is obtained similarly from $W(p)$.  For
entirely spatial $p$,
  \eqn{WZZ}{\eqalign{
   W_{00}(p) &= {i \over \vec{p}^2 + m_\gamma^2}  \cr
   W(p) &= -{i \over \vec{p}^2 + m_\phi^2} \,.
  }}
 Thus we see that the potential energies from gauge bosons and from
scalars have the same functional form, up to a sign.  The repulsion
due to the gauge bosons dominates when $m_\gamma < m_\phi$.  Some
subtleties on the normalization of the potential will be discussed in
section~\ref{GaugeField}.  In section~\ref{BulkForce} we will argue
that the bulk contribution is negligible.

The only other ingredient necessary to compute the force is the strength 
of the coupling between the gauge field and the electrons. In 
section~\ref{GaugeField} we shall show that the electron couples with a 
strength $q$ given by 
\eqn{coplstr}{
q = \sqrt{ { 1 \over 2 \nu l_s} \left( {\pi^2  \over N }\right)^{1/3}}.
}
The momentum space potential contributed by the gauge field is 
  \eqn{MomPot}{
   V_\gamma(\vec{p}) = q^2  W_{00}(\vec{p}) = 
    q^2 {i \over \vec{p}^2 + m_{\gamma}^2} \,.
  }
 Fourier transforming back to position space gives
  \eqn{gbpot}{
   V_\gamma(r) = 2 \pi q^2 K_0(m_{\gamma} r) \,.
  }
 Recalling that the scalar contribution has the same functional form
as $V_\gamma$, and that the net force should vanish in the $r \to 0$
limit because the BPS property is asymptotically recovered at short
distances, we conclude that the total tree-level potential from the
D2-brane gauge theory is
  \eqn{TotPot}{
   V_{\rm bdy}(r) = 2\pi q^2 \left[ K_0(m_\gamma r) - K_0(m_\phi r) 
    \right] \,.
  }
 The potential scales with $N$ and $\nu$ like $q^2 \sim {1 \over \nu
N^{1/3}}$.

\subsection{The gauge field}
\label{GaugeField}
In \cite{bbst}, the effect of six-branes in the set-up was modeled by 
replacing them by their gravitational background, which we reproduce here for 
convenience. The spacetime metric (in rescaled coordinates) for $K$ D6-branes
is 
\eqn{sixbranebg}{
ds^2 = \sqrt{{\rho \over l_s} } \left( d \tau^2 - d \tilde{y}^a  d \tilde{y}^a
\right) - \sqrt{{ l_s \over \rho}} \left(d \rho^2 + \rho^2 d \Omega_2^2 \right)
}
 and the background dilaton is 
\eqn{bgdilaton}{
g_s^2 e^{2 \Phi} = {4 \over K^2 } \left( { \rho \over l_s} \right)^{3/2}
  \,.
}
 It was also shown that a spherical D2-brane with $N$-units of magnetic flux 
is stable at a radius given by 
\eqn{stabrad}{ 
\rho_* = { (\pi N )^{2/3} \over 2 } \; l_s
} 
 Since the gauge theory lives on the world-volume of the D2-brane, we can infer
that the Yang-Mills coupling of the theory is given as 
\eqn{ymcoupl}{
g_{YM}^2 l_s = g_s e^{\Phi}|_{\rho_*} = 2^{{ 1 \over 4}} { \sqrt{ \pi N} 
\over K}.
}

The authors of \cite{bbst} give an open string metric, Eq.~(5.15) to
be precise, which is computed using the standard Seiberg-Witten
prescription \cite{swNC} using a flat closed string metric and a
B-field of appropriate strength.  The D2-brane unfortunately does not
live in flat space.  The induced metric in the static gauge is 
 \eqn{indmet}{
ds^2_{\rm ind} = {(\pi N )^{1/3} \over \sqrt{2}} d\tau^2 - 
{(\pi N)\over 2 \sqrt{2}} l_s^2 (d\theta^2 + 
\sin^2 \theta \; d\phi^2 ) \,.
}
\noindent
Using the B-field to be $ B_{\theta\phi} = { N \over 2 } \sin \theta$, 
we can evaluate the correct open string metric as seen by the D2-brane 
world volume theory as
\eqn{opstmet}{
ds^2_{\rm open}  =  {(\pi N)^{1/3} \over \sqrt{2} } 
\left( d\tau^2 - {  9 l_s^2 \over 2  }
(\pi N)^{2/3} (d\theta^2 + \sin^2 \theta \; d \phi^2 ) \right) \,.
}
\noindent
It is this metric that appears in the non-commutative gauge boson kinetic 
term. So to get the right normalizations for the gauge bosons,
start from the action 
\eqn{gakin}{
S_{\rm gauge} = - {1 \over 2 g_{YM}^2} \int d^3 \xi \sqrt{G_{\rm open}} \;
G_{\rm open}^{\mu \rho} G_{\rm open}^{\nu \sigma} F_{ \mu \nu} F_{\rho \sigma} \,.
}
\noindent
To ensure that we write the scalar and the gauge boson action in terms of the 
same time coordinate, let us conformally rescale the metric by writing the 
action in terms of a metric we call $\tilde{G}_{\rm open}$. Since we would like to 
put our action in canonical form as in \LowS\ we would need to rescale the 
gauge fields to achieve this end. Defining
$ \alpha = {(\pi N)^{1/3} \over \sqrt{2}} $, we want  
$\alpha \tilde{G}_{\rm open} =  G_{\rm open}$, implying 
$\sqrt{G_{\rm open}} \;
G_{\rm open}^{\mu \rho} G_{\rm open}^{\nu \sigma} = {1 \over \sqrt{\alpha}}
\sqrt{\tilde{G}_{\rm open}} \;
\tilde{G}_{\rm open}^{\mu \rho} \tilde{G}_{\rm open}^{\nu \sigma}.$
Hence we can cast the gauge boson kinetic term in the canonical form by 
defining, $A  =  \alpha^{1/4} \sqrt{{g_{YM}^2 \over 2 }} \tilde{A}$. 
The gauge boson mass term is
\eqn{gbmass}{\eqalign{
{ K \over 4 \pi N } \int d^3 \xi 
\epsilon^{\mu \nu \sigma} A_{\mu} \partial_\nu A_{\sigma} 
& = { K \over 4 \pi N } \int d^3 \xi 
\sqrt{\alpha} {g_{YM}^2 \over 2} \; \epsilon^{\mu \nu \sigma} \;
 \tilde{A}_{\mu} \partial_\nu \tilde{A}_{\sigma} \cr
& =\int d^3 \xi \; { m_{\gamma} \over 2} \; \epsilon^{\mu \nu \sigma} 
\tilde{A}_{\mu} \partial_\nu \tilde{A}_{\sigma}
}}
\noindent
with 
$$m_{\gamma} = { K  \over 4 \pi N} g_{YM}^2 \sqrt{\alpha} = 
{1 \over 4} { 1 \over (\pi N )^{1/3} l_s} \,.$$ 
\noindent
In the above series of manipulations we have taken cognizance of the fact that
the Chern-Simons term is topological and hence will remain unaffected by the 
conformal rescaling of the metric. Note that the Compton wavelength of the 
photon is indeed of the same order as the size of the sphere measured in 
units prescribed by the metric $\tilde{G}_{\rm open}$. 

One other ingredient that will be necessary is a proper normalization of the 
coupling of the ``electrons'' to the gauge field. This normalization can 
be fixed by comparing the coupling to the chemical potential term, for 
the system is constrained to have exactly $K$ ``electrons.'' Writing the action
in terms of the rescaled variables introduced above, we have 
\eqn{gbcoupl}{
S_{\rm coupling} = \sum_{i = 1}^{K}\; { 1 \over V}  \int d^3 \xi 
\alpha^{1/4} \sqrt{{g_{YM}^2 \over 2 }} \tilde{A}  + \int d^3 \xi \;
\tilde{A}_0 \tilde{J}^0}
\noindent
with $ \tilde{J}^0 = { K \over V } \alpha^{1/4} \sqrt{{g_{YM}^2 \over 2 }}$.
Varying the above with respect to $\tilde{A}$ we see that the chemical 
potential term is saturated by the presence of $ K $ electrons.
The main point of this is that an electron couples to the 
gauge field $\tilde{A}$ with strength 
$q = \alpha^{1/4} \sqrt{{g_{YM}^2 \over 2}}$, as promised in \coplstr.

\subsection{The scalar action}
\label{ScalarField}

The DBI action for the 2-brane (treated as a probe) in the near-horizon 
geometry of the D6-branes was used to compute the potential of the radial mode
scalar, and to show that there is indeed a radius wherein the D2-D0 
bound state could be stabilized. Indeed the same approach can be extended to 
compute the scalar kinetic terms, a necessary ingredient in determining 
the mass of the radial mode. 

Choosing to work in static gauge with coordinates
\eqn{static}{
\xi^0 = \tau , \;\; \xi^1 = \theta, \;\; \xi^2 = \phi \,, 
} 
\noindent 
the induced metric, given the spacetime metric \sixbranebg, turns out to be 
\eqn{indmetTwo}{\eqalign{
G_{00} &= \sqrt{{\rho \over l_s}} - \sqrt{{l_s \over \rho}} \left( 
{\partial \rho \over \partial \tau} \right)^2 \cr
G_{11} &= - \sqrt{{\rho^3  l_s}} - \sqrt{{l_s \over \rho}} \left( 
{\partial \rho \over \partial \theta} \right)^2 \cr
G_{22} &= \sqrt{{\rho^3 l_s}} \sin^2 \theta  - \sqrt{{l_s \over \rho}} \left( 
{\partial \rho \over \partial \phi } \right)^2. 
}}
\noindent
In addition we have the world-volume field strength turned on, 
\eqn{fieldstr}{
F_{12} = {N \over 2 } \sin \theta \,.
}
\noindent
Plug all of this into the DBI action:
\eqn{dbi}{\eqalign{
L_{DBI} &= - { 1 \over 4 \pi^2 g_s l_s^3 } \int \; d\tau d\theta d\phi \;
e^{-\Phi} \det[G_{ab} + 2 \pi l_s^2 F_{ab}]^{{1 \over 2}} \cr
        & = {K \over 8 \pi^2 l_s^2} \; \int d^3 \xi \; \rho
\sqrt{1 + {N^2 \pi^2 l_s^3 \over \rho^3}} 
\left\{
-1 +
{1 \over 2}  {l_s \over \rho}
\left({\partial \rho \over \partial \tau} \right)^2 
- {1 \over 2} \left( \nabla_T \rho \right)^2
\left({1 \over \rho^2}\; { 1 \over 1 + {N^2 \pi^2 l_s^3 \over \rho^3}}  \right)
\right\} 
}}
\noindent
The first term in the action (the $-1$ part) is just the potential term that
was evaluated in \cite{bbst}.
Expanding about the critical point $ \rho = \rho_*$ we get
\eqn{scaact}{\eqalign{
S_{\rm scalar} &= 
 \int d^3 \xi \; \left\{ { 3 K \over 8 \pi^2 l_s} \left[ { 1 \over 2} 
\left({\partial \rho \over \partial \tau} \right)^2 - {2 \over 9 l_s^2 
(\pi N)^{2/3} } {1 \over 2 } \left( \nabla_T \rho \right)^2 \right]
- {2 \over 3 \pi^2 l_s^3} {K \over (N\pi)^{2/3}} 
\rho^2 \right\} \cr
&= \int d^3 \xi \; \sqrt{\tilde{G}_{\rm open}} \; \left[ { 1 \over 2}
\left({\partial \phi \over \partial \tau} \right)^2  - { 1 \over 2}
{2 \over 9 l_s^2 (\pi N)^{2/3} }
\left( \nabla_T \phi \right)^2 - 
{ 16 \over 9 l_s^2} {1 \over (\pi N)^{2/3}} \phi^2 \right] \,.
}}
\noindent
We have rescaled the scalar $\rho$ and written the action in terms of a 
new scalar $\phi$, so that the metric seen by the scalars is also 
$\tilde{G}_{\rm open}$. This gives
$$ m_{\phi}^2 = { 32 \over 9 l_s^2} {1 \over (\pi N)^{2/3}} \,.$$

Hence we find that $m_{\phi}$ and $m_{\gamma}$ are of the same order
in $N$, but that the scalar is heavier by a numerical factor of
$16\sqrt{2}/3$.

\subsection{Estimating the bulk force}
\label{BulkForce}

It is difficult to compute the force between a pair of strings due to
the exchange of massless string modes: the strings are finite in
length, and the background is non-trivial.  However, by making some
reasonable assumptions, we can estimate the potential arising from
closed string exchange.  Let $\vartheta$ be the angle between a pair
of strings.  Our first assumption is that the potential has the form
  \eqn{VBulk}{
   V_{\rm bulk}(\vartheta) = V_0 (1-\cos\vartheta) \,.
  }
 While this may not be exactly right, it seems very likely to be close
enough for our purposes.  More specifically, the property which we
expect the true $V_{\rm bulk}(\vartheta)$ to share with \VBulk\ is
that it has only one scale: the maximum value of $V_{\rm
bulk}(\vartheta)$ is of the same order of magnitude as the second
derivative at $\vartheta = 0$ (this is the ``plausible technical
assumption'' mentioned in the introduction).  A mild singularity at
$\vartheta = 0$, such as a $\vartheta^2 \log\vartheta$ term, would not
affect our conclusions.  A sharper singularity at $\vartheta = 0$
would be unexpected since $\vartheta = 0$ is where a no-force
condition is restored.

Second, we assume that $V_0$ may be estimated as the magnitude of the
gravitational potential energy experienced by two point masses in flat
ten-dimensional space, separated by the same distance as the endpoints
of the strings at angle $\vartheta = \pi/2$, and having the same mass
as the strings.  This assumption is safe as long as there isn't a
strong force coming from the region very close to the D6-brane.  The
mass of the strings is\footnote{We are defining mass by integrating
the Nambu-Goto action of the string over the spatial coordinate:
$S_{NG} = -{1 \over 2\pi\alpha'} \int d^2 \sigma \, \sqrt{\det
G_{\mu\nu}} = -m_{\rm str} \int d\tau$ for a static string.  This results in a
slightly different answer from \cite{bbst}, but we believe our
approach is the correct one for our purposes.}
  \eqn{StringMass}{
   m_{\rm str} = {1 \over 2\pi l_s} {\rho_* \over l_s} \sim {N^{2/3} \over l_s}
  }
 and they are separated by a distance 
  \eqn{SeparationLength}{
   L \sim \rho_* \left( {l_s \over \rho_*} \right)^{1/4} 
     \sim l_s \sqrt{N} \,.
  }
 Newton's constant is $G_N = \left. g_s^2 e^{2\phi} \right|_{\rho_*}
\sim l_s^8 N/K^2$.  The gravitational potential energy has magnitude
  \eqn{GravPot}{
   V_0 = {G_N m_{\rm str}^2 \over L^7} \sim {1 \over l_s} 
    {1 \over \nu^2 N^{19/6}} \,.
  }
 Clearly, in a $N \to \infty$ limit with $\nu$ held fixed, $V_0$
scales to zero much faster than the magnitude of the potential induced
by D2-brane effects (see the text following \TotPot).  Thus we are
justified in asserting that bulk effects are negligible.  This is
gratifying because it verifies that we are working in a decoupling
limit, where gravitational effects are much weaker than the open string
effects on the D2-brane world-volume.

\section{Quantum Hall fluid or Wigner crystal?}
\label{WhichPhase}

In general it is difficult to be sure whether a particular Hamiltonian
will or won't lead to quantum Hall behavior without performing some
diagonalization or robust variational calculation.  However a figure
of merit which serves as a useful guide to the physics is a typical
energy of interactions divided by the cyclotron gap,
  \eqn{EtaDef}{
   \eta = {V_{\rm typ} \over \omega_c} \,.
  } 
 Physically, the smallness of this ratio is a measure of the validity
of projecting the system to the lowest Landau level and treating the
interactions perturbatively.  If $\eta$ is small, a quantum Hall
ground state is clearly favored.  If $\eta$ is very large, then there
is no reason for the ground state to be close to a combination of
lowest Landau level wave-functions; instead one may expect a Wigner
crystal to win out energetically.

To compute the cyclotron gap, one should in principle start from the
string worldsheet and compute the mode of excitation corresponding to
cyclotron motion.\footnote{We thank L.~Susskind and N.~Toumbas for
pointing out to us that our original estimate of $\omega_c$ was
parametrically smaller than the correct answer, leading us to the
incorrect conclusion that the Wigner crystal was favored over the
quantum Hall state.  An approximate worldsheet calculation was also
supplied to us by L.~Susskind, which leads to an answer similar to the
computation presented here.}  However, because this is the lowest
excitation mode available to the string, it is a fair approximation to
say that the string moves rigidly.  Thus it suffices, at least for the
purposes of estimating $\omega_c$ up to factors of order unity, to
keep track only of the dynamics of the end of the string.  For this
purpose we need the action
  \eqn{StringLag}{
   S = -m \int ds + q\int \tilde{A} \,.
  }
 The line element $ds$ is defined by $ds^2 = \tilde{G}^{\rm
open}_{\mu\nu} d\xi^\mu d\xi^\nu$, so that for a static string, $ds =
d\tau$.  The mass $m$ is $m_{\rm str}$ computed in \StringMass.  Let
us assume that the string end is near the equator, $\theta = \pi/2$,
of the ${\bf S}^2$, and choose local coordinates $x_1$ and $x_2$ for
the position of the string end such that $dx_1 = d\theta$ and $dx_2 =
d\phi$.  Then, setting $l_s = 1$ and dropping up to factors of order
unity, $q\tilde{F}_{12} = QF_{12} \sim N$, $m \sim N^{2/3}$, and $ds =
d\tau \sqrt{1 - N^{2/3} \dot{x}_\alpha^2}$.  Making the
non-relativistic approximation where the square root in the last
expression for $ds$ is expanded out to leading order, one finds from
\StringLag\ the equations for cyclotron motion with
  \eqn{OmegaC}{
   \omega_c = {QF_{12} \over N^{2/3} m} \sim N^{-1/3} \,.
  }

In real quantum Hall systems, typical interaction energies are
computed as $e^2/\ell$ where $\ell$ is the average separation between
nearest neighbors.  Here the potential behaves as $q^2 r^2 \log
m_\gamma r$ for small separations $r$, so we might conclude that
interactions are very weak indeed, and that a quantum Hall ground
state is favored.  This is overly optimistic, since the forces between
non-nearest-neighbors have not been accounted for.  To obtain a more
conservative estimate of the typical interaction energy, let us assume
that the electrons are approximately evenly spaced on the sphere, and
see what energy it would take to move one electron to the location of
its nearest neighbor.  The magnitude of this energy may be estimated
as $V_{\rm typ} \sim q^2 \sim 1/(\nu N^{1/3})$ in string units.  (One
way to get at this is to replace the sphere by a circular patch with a
uniform density of electrons, and then compare the potential energy of
an electron at the center of the circle to one slightly displaced from
the center).  Thus the figure of merit turns out to be $\eta \sim
1/\nu$.

Because we find no parametric dependence of $\eta$ on $N$, we cannot
with confidence claim that the ground state at moderate values of
$\nu$ will be a quantum Hall liquid or a Wigner crystal in the
thermodynamic limit where $N \to \infty$.  Real quantum Hall systems
in fact have $\eta$ considerably larger than $1$.  In fact, as
remarked in section~\ref{CondMat}, such systems have a Wigner crystal
phase for small $\nu$ ($\nu \lsim 1/7$), a quantum Hall phase at
slightly higher $\nu$ (for instance, $\nu \approx 1/3$), and yet other
phases, like stripes, at larger $\nu$ (like $\nu \approx 9/2$).  The
D-brane system should exhibit a Wigner crystal phase at small $\nu$,
though for different reasons than real quantum Hall systems: for the
D-brane system, $\eta \sim 1/\nu$, so the repulsive interactions
become strong as $\nu \to 0$.  This seems backwards in comparison to
real quantum Hall systems, where $\nu \to 0$ corresponds to extremely
strong magnetic field.  The D-brane system is different in that the
magnetic flux per unit area is essentially constant.  The ${\bf S}^2$
adjusts its size to make it so.  Small $\nu$ pushes up the gauge
coupling, and with it the strength of the repulsive interactions.

On the other hand, if $\nu$ is too large, then string excitations are
energetically available which reverse the statistics of the electrons.
Different excited string states should be viewed as separate species
of particles, but with the same electric charge.  It's not clear what
the ground state will be in this case.  Stripes with integer filling
fractions of each species is perhaps a reasonable guess---but the
system is complicated, as can be seen from the breakdown of the
arguments in \cite{bbst} that quasi-particle energies are smaller than
other energy scales in the system.

\begin{figure}[h]
   \centerline{\psfig{figure=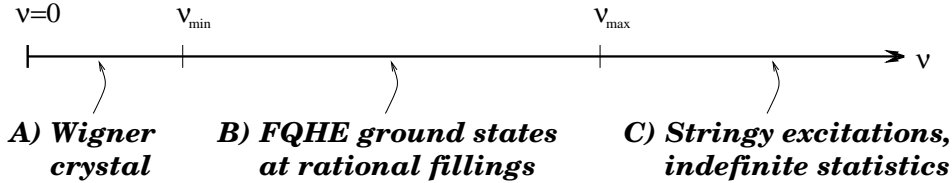,width=5in}}
   \caption{Expected ``phase diagram'' of the quantum Hall soliton.
It would take numerics to figure out the value of $\nu_{\rm min}$.  If
the electrons are fermions in their ground state, a reasonable guess
is $\nu_{\rm max} = 1$.  If they are bosons, multiple occupancy of
lowest Landau level states may push $\nu_{\rm max}$ somewhat higher.}
   \label{figB}
  \end{figure}

For an intermediate range of $\nu$, where interactions are under
reasonable control but only the lowest string mode is available, one
may hope that a conventional quantum Hall liquid is indeed the ground
state of the system (see figure~\ref{figB}).
 This expectation should be born out by numerics, comparing a
Laughlin-like wavefunction to a Wigner crystal for various values of
$\nu$.  It would be nice to contrive a D-brane set-up where
interactions can be made parametrically weak, so that fractional
quantum states are clear winners over any other state of the system
for some range of $\nu$.  To do this, one might for instance try to
lower the mass of the electrons (and thereby raise the cyclotron gap)
by having the strings end not on the D6-branes but on some other brane
closer to the D2-brane.  An anti-D2-brane concentric with the D2-brane
might approximately fit the bill.  Precisely this possibility was
discussed in \cite{bbst}, and it was found that there was no energy
barrier toward creating such an anti-D2-brane in the strict
near-horizon limit for the D6-brane; but restoring the one to the
harmonic function throws up a slight potential gradient preventing it.
On top of this there is a brane-anti-brane attraction.  Although our
investigation has not been detailed, it seems that the only stable (or
meta-stable) equilibrium point is the one with no anti-D2-brane: in
particular, the ``phenomenologically'' attractive configuration where
the anti-D2-brane sits very close to the D2-D0 bound state appears to
be unstable.  More elaborate brane configurations worthy of
consideration include intersecting or nearly-intersecting branes with
the electrons arising from short strings between two nearby branes.

\section*{Acknowledgments}

We would like to thank J.~Maldacena, I.~Klebanov, L.~Susskind,
N.~Toumbas, A.~Vishwanath and especially S.~Sondhi for useful
discussions.  The work of S.S.G.\ was supported in part by DOE
grant~DE-FG02-91ER40671, and by a DOE Outstanding Junior Investigator
award.

\bibliography{qhe}
\bibliographystyle{ssg}

\end{document}